\documentclass[aps,prl,preprint,groupedaddress,nofootinbib]{revtex4-2}
\usepackage{graphicx}
\usepackage{dcolumn}%
\usepackage{bm}
\usepackage{mathtools}
\usepackage[version=3]{mhchem}
\usepackage[left=1.5cm, right=1.5cm, top=1.785cm, bottom=2.0cm]{geometry}
\usepackage{balance}
\usepackage{amssymb} 
\usepackage{amsmath} 
\usepackage{mathptmx}
\usepackage{lastpage}
\usepackage{color}
\usepackage{soul}
\usepackage{comment}
\usepackage{braket}
\usepackage[format=plain,justification=justified,singlelinecheck=false,font={stretch=1.125,small,sf},labelfont=bf,labelsep=space]{caption}
\usepackage{float}
\usepackage{fancyhdr}
\usepackage{fnpos}
\usepackage[english]{babel}
\addto{\captionsenglish}{
 
}
\usepackage{array}
\usepackage{droidsans}
\usepackage{charter}
\usepackage[T1]{fontenc}
\usepackage[usenames,dvipsnames]{xcolor}
\usepackage{setspace}
\usepackage[compact]{titlesec}
\usepackage{hyperref}
\hypersetup{
  colorlinks   = true, 
  urlcolor     = blue, 
  linkcolor    = blue, 
  citecolor   = red 
}

\usepackage{epstopdf} 

\usepackage[normalem]{ulem}
\usepackage{framed}
\makeatletter
\newcommand\footnoteref[1]{\protected@xdef\@thefnmark{\ref{#1}}\@footnotemark}
\makeatother
\newcommand{\Fig}[1]{Fig.\,{#1}}

\newcommand{\del}[1]{\frac{\partial}{\partial #1}} 
 
\newcommand{\delc}[2]{\frac{\partial #1}{\partial #2}}

\newtheorem{lemma}{Lemma}

\begin{document}

\title{Geometry of chiral temporal structures II: The formalism}

\author{Aycke Roos$^{1}$, Pablo M. Maier$^{1}$, Andres F. Ordonez$^{2,3}$ and Olga Smirnova$^{1,4,5}$}

\affiliation{
$^1$Max-Born-Institut, Max-Born-Str. 2A, 12489 Berlin, Germany\\
$^2$Department of Physics, Imperial College London, SW7 2BW London, United	Kingdom\\
$^3$Department of Physics, Freie Universit\"at Berlin, 14195 Berlin, Germany\\
$^4$Technische Universit\"at Berlin, 10623 Berlin, Germany\\
$^5$Technion - Israel Institute of Technology, Haifa, Israel}

\date{\today}
\begin{abstract}
We develop a mathematical formalism underlying the emergence of  enantio-sensitive molecular orientation due to photoionization or photoexitation of chiral molecules. We consider geometric quantities such as the Berry connection and Berry curvature in light-driven chiral electronic states in the space of complex light polarization vectors. The parametric dependence of the light-driven electronic wavefunction on such vectors emerges due to various possible mutual orientations between the laser field and a chiral molecule. Using the tools of differential geometry we show how the enantio-sensitive observables emerge from the geometry of the molecular response in such spaces.
\end{abstract}

\maketitle
\section{1. Introduction}
Light-induced ultrafast electronic currents in chiral molecules \cite{Beaulieu2018PXCD, wanie2024capturing} can trace temporal chiral structures even when the driving laser field itself is achiral. A chiral temporal structure is a trajectory that can not be superimposed on its mirror image \cite{ayuso2022ultrafast}.
Inspired by the emergence of geometric properties in the photoionization of chiral molecules \cite{ordonez2022geometric, ordonez2023geometric} we develop a unified approach to understanding and controlling enantio-sensitive geometric observables based on the analysis of temporal shapes of chiral electron currents. Our framework is naturally geometric, grounded in the formalism of fiber bundle theory \cite{husemoller1966fibre, bouchiat1988non} over complex polarization spaces \cite{bliokh2019geometric}. These complex vector spaces describe the polarization of the driving laser field and allow a differential-geometric description of enantio-sensitive phenomena. In this setting, the wavefunction defines a section of a fiber bundle whose Berry connection and Berry curvature yield physical observables.

Our generalized approach reveals strategies for controlling geometric quantities such as the Berry curvature and Berry connection through multiphoton processes driven by polarization-shaped pulses targeting either geometries of bounds states via photoexcitation or continuum states via photoionization or both. These opportunities are described in our companion paper \cite{ordonez2024geometrytemporalchiralstructures}.
Here we focus on using the tools of differential geometry \cite{nakahara2018geometry, bohm2013geometric} to derive the Berry connection and the Berry curvature in complex vector spaces to predict and analyze new enantio-sensitive observables of geometric origin.

Choosing the electric field vector space as the parameter space is a choice that emphasizes physical intuition, though the molecule’s orientation remains the fundamental control parameter. This choice enables a clear geometric formulation of enantio-sensitive responses, and allows us to define a Berry connection and Berry curvature in the space of light polarization vectors. The orientation-dependence of the wavefunction originates entirely from the dipole interaction. In the molecular frame, the electric field polarization vector can be seen as a tangent vector being parallely transported over the space of molecular orientations as the molecule rotates. For a circular field polarization this space reduces to the sphere $S^2$. The polarization vector can be naturally viewed as an element of the complexified tangent space to the sphere, and thinking of the Berry connection and Berry curvature in terms of this complex tangent vector makes the geometric nature of the phase explicit. The geometric phase of the wave function can thus be related to the phase acquired by a complex vector undergoing parallel transport along the sphere. The $U(1)$ phase reflects the rotation of this complex tangent vector within the tangent plane.

In the following sections, we develop this geometric framework explicitly and show how key physical observables—such as molecular orientation or circular dichroism—are encoded in the Berry curvature and Berry connection defined over these complex vector spaces.



\section{2. Key geometric quantities relevant for enantio-sensitive observables}

As introduced in the companion paper \cite{ordonez2024geometrytemporalchiralstructures} our analysis of the geometric structure of the manifold of temporal shapes in the space of electric field polarization vectors $\bm e$ reveals two key quantities. The first one is the geometric phase \cite{moore1992adiabatic}:
\begin{align} 
  S=\oint i\langle \psi(\bm{e})|\nabla_{\bm{e}}\psi(\bm{e})\rangle \cdot d\bm{e}=\oint \bm A(\bm{e})\cdot d\bm{e},
 \label{eq:S} 
\end{align}
revealing  the Berry connection vector defined as:
\begin{equation} \bm A(\bm{e}) = i\langle \psi(\bm{e}) | \nabla_{\bm{e}} \psi(\bm{e}) \rangle. \label{eq:connection_t} \end{equation}
This vector quantifies the sensitivity of the electronic wavefunction $\psi(t, \bm{e}(\rho), \bm{r})$ to changes in the orientation of the laser field polarization $\bm{e}(\rho)$ in the molecular frame and $\rho$ describes Euler angles quantifying the orientation of molecular frame with respect to the laboratory frame. The wavefunction $\psi(t, \bm{e}(\rho), \bm{r})$, expressed in the molecular frame, is the solution of the time-dependent Schrödinger equation for a fixed relative orientation $\rho$ between the molecule and the laser field:
\begin{equation} \label{eq:TDSE_local} i \del{t} \psi(\bm{r}, \rho, t) = \left[ H_{el} + \bm{r} \cdot \bm{E}(\rho, t) \right] \psi(\bm{r}, \rho, t). \end{equation}
The vector $\bm A(\bm{e})$ "connects" wave-functions at $\psi(\bm{r}, \bm{e}(\rho), t)$ and $\psi(\bm{r}, \bm{e}(\rho+d\rho), t)$.  The second relevant geometric quantity is a pseudovector, defined as the curl of the Berry connection in the space of electric field polarization vectors:
\begin{equation} \bm{\Omega}_a = \nabla_{\bm{e}} \times \bm{A} = i \langle \nabla_{\bm{e}} \psi(\bm{e}) | \times | \nabla_{\bm{e}} \psi(\bm{e}) \rangle. \label{eq:curvature_def} \end{equation}
Equations~(\ref{eq:connection_t}, \ref{eq:curvature_def}) highlight the dynamical, laser-driven origin of these geometric quantities.

The pseudovector $\bm{\Omega}_a$ identifies a well-defined direction in the molecular frame. In our companion paper \cite{ordonez2024geometrytemporalchiralstructures} we show that the direction of this vector determines the orientation of a molecular cation upon two-photon ionization of a randomly oriented ensemble of molecules and thus it is a directly accessible experimental observable. As such, it deserves a dedicated analysis that we perform in this paper (Sections 3,4,5). 
In a real-valued, three-dimensional parameter space the curl of the vector potential, i.e. the pseudovector $\bm{\Omega}$ corresponds to the Berry curvature. In the following, we show that in complex vector spaces the Berry curvature tensor gets additional components. 

\section{3. The Emergence of the Berry Curvature in One-Photon Ionization} \label{Sec3} 
Let us consider a one-photon ionization from an initial state $\psi_0$ by circularly polarized light to follow the emergence of the Berry curvature. We can use perturbation theory to calculate $|\psi(\bm{e})\rangle$:  
\begin{align} 
 \label{eq:psi} 
\psi & =\psi_{0}+ \int\mathrm{d}\Theta_{\bm{k}}a_{\bm{k}}(\bm{e})\psi_{\bm{k}}
\end{align} 
and thus evaluate the Berry connection explicitly.
Here $\mathrm{d}\Theta_{\bm{k}}=\sin\theta_{k}\mathrm{d}\theta_k\mathrm{d}\phi_k$ and $\theta_{k}$, $\phi_{k}$ are angles characterizing the direction of the photoelectron momentum $\bm{k}$ for fixed $k=|\bm{k}|$, and $a_{\bm{k}}=-\widetilde{E}_{\omega_k}\left(\bm{D}\cdot\hat{\bm{e}}\right)$ are the amplitudes of the continuum states and $\widetilde{E}_{\omega_k}$ is the Fourier transform of the electric field amplitude at the transition frequency. The polarization of the laser field in the molecular frame is characterized by a complex unit vector $\hat{\bm{e}} =\frac{1}{\sqrt{2}}\left(\hat{\bm{\theta}}+i\sigma\hat{\bm{\phi}}\right)$ that encodes the direction of field rotation $\sigma=\pm1$. Each component $e^j$ is a function defined on the 2-sphere (see Figure 1).
The Berry connection becomes: 
\begin{align}
  \mathcal{A}=i\langle\psi|d\psi\rangle=i\int\mathrm{d}\Theta_{\bm{k}}(\bm{D} \cdot \hat{\bm{e}})^{*}(\bm{D} \cdot d\hat{\bm{e}}).
\end{align} 
We can expand the exterior derivative of the vector $\hat{\bm{e}}$ in terms of the unit Cartesian basis vectors $\hat{\bm{j}}$ in the molecular frame,
\begin{align}
  d\hat{\bm{e}} = (de^j) \hat{\bm{j}}. 
\end{align}

In this framework, the Berry curvature is the exterior derivative of the Berry connection $\varOmega=d\mathcal{A}$:
\begin{align}
 \Omega = i|\widetilde{E}_{\omega_k}|^2\int\mathrm{d}\Theta_{\bm{k}}\, D_i^* D_j \, de^{*i} \wedge de^j.
\label{eq:curvature_full_1}
\end{align}
Note that in Eq.\eqref{eq:curvature_full_1} the Berry curvature is naturally factorized into a molecular tensor -- the dipole tensor $\bm{\Omega}^{ij} = i\int\mathrm{d}\Theta_{\bm{k}}\, D_i^* D_j$ -- and light polarization two form $de^{*i} \wedge de^j$. Since the dipole tensor is anti-hermitian it may be partitioned into a symmetric (imaginary) and anti-symmetric (real) matrix, 
\begin{align}
\bm{\Omega}^{ij}_s &\equiv \frac{i}{2} |\widetilde{E}_{\omega_k}|^2\int\mathrm{d}\Theta_{\bm{k}}\left( D_i^* D_j + D_j^* D_i \right), \\
\bm{\Omega}^{ij}_a &\equiv \frac{i}{2}|\widetilde{E}_{\omega_k}|^2\int\mathrm{d}\Theta_{\bm{k}} \left( D_i^* D_j - D_j^* D_i \right), 
\end{align}
such that $\bm{\Omega}^{ij}=\bm{\Omega}_{s}^{ij} +\bm{\Omega}_{a}^{ij}$. The real part of the dipole tensor is vectorizable and can be expressed in terms of a cross product, 
\begin{align}
\bm{\Omega}^{ij}_- &= \frac{i}{2} |\widetilde{E}_{\omega_k}|^2\epsilon^{lij}\int\mathrm{d}\Theta_{\bm{k}} (\bm{D}^* \times \bm{D})_l,
\label{eq:S_a}
\end{align}
where \( \epsilon^{lij} \) is the Levi-Civita symbol.

On the basis of Eq. \eqref{eq:S_a}, we obtain the expression for the antisymetric part of the Berry curvature:
\begin{align}
\Omega_{a} 
= \bm{\Omega}_{a}^{ij} de^{*i} \wedge de^j
= i |\widetilde{E}_{\omega_k}|^2\epsilon^{lij} \int\mathrm{d}\Theta_{\bm{k}}(\bm{D}^* \times \bm{D})_l de^{*i} \wedge de^j. 
\end{align} 
It is now natural to define the \emph{polarization Berry curvature 2-forms}:
\begin{align}
d\xi^l \equiv \frac{1}{2}\epsilon^{lij} \, de^{*i} \wedge de^j,
\end{align} 
and their molecular counterparts, 
\begin{align}
\bm{\Omega}_{a} \equiv i|\widetilde{E}_{\omega_k}|^2\int\mathrm{d}\Theta_{\bm{k}} \bm{D}^* \times \bm{D},
\end{align} 
such that the orbital antisymmetric contribution to the Berry curvature becomes 
\begin{align}
\Omega_{a} = \bm{\Omega}_{a} \cdot \bm{d\xi},
\end{align}
where we introduce the term orbital antisymmetric to distinguish it from antisymmetrization with respect to the spin degree of freedom, which is not considered here. We deduce a very important result: the orbital antisymmetric component of the Berry curvature tensor $\bm{\Omega}$ reduces, when paired with the polarization differential $de^{*i}\wedge de^j$, to the scalar product of a molecular pseudovector and a surface element $\bm{d\xi}$. This implies that $\bm{\Omega}_{a}$ specifies a well-defined and molecular-specific direction in the molecular frame. Its physical meaning has been identified in our earlier work \cite{ordonez2023geometric}. Specifically, $\bm{\Omega}_{a}$ is equivalent to the net propensity field \cite{ordonez2022geometric, ordonez2019propensity2, ordonez2023geometric, ordonez_propensity_2019, ordonez_propensity_2022}. Its projection on photoelectron spin quantifies circular dichroism in one-photon ionization \cite{ordonez2023geometric}. Its direction quantifies enantio-sensitive orientation of molecular cations upon photoionization from a current carrying state \cite{ordonez2023geometric}. This pseudovector also plays an important role in spin-orientation locking in chiral molecules.
However, the symmetric counterpart of the Berry curvature can not be reduced to a cross product, 
\begin{align}
\Omega_{s} = \bm{\Omega}^{ij}_{s}de^{*i} \wedge de^j.
\end{align}
Note that the orbital symmetric and orbital antisymmetric parts of the Berry curvature arise from the following components of the molecular dipole tensor, \begin{align}
\bm{\Omega}_{s} &= |\widetilde{E}_{\omega_k}|^2\int\mathrm{d}\Theta_{\bm{k}}\mathrm{Im}\{iD_i^{*}D_j\}=\mathrm{Re} \braket{\partial_{e^i}\psi|\partial_{e^j}\psi} \\ 
\bm{\Omega}_{a} &= |\widetilde{E}_{\omega_k}|^2\int\mathrm{d}\Theta_{\bm{k}}\mathrm{Re}\{iD_i^{*}D_j\}=-\mathrm{Im} \braket{\partial_{e^i}\psi|\partial_{e^j}\psi}\label{eq:Omega_a}. \end{align} 

Usually only the imaginary part $\mathrm{Im}\braket{\partial_{e^i}\psi|\partial_{e^j}\psi}$ of a tensor $\braket{\partial_{e^i}\psi|\partial_{e^j}\psi}$ contributes to the Berry curvature (see e.g. \cite{cheng2013quantumgeometrictensorfubinistudy}), however in our case $\mathrm{Re}\braket{\partial_{e^i}\psi|\partial_{e^j}\psi}$ also contributes due to the fact that our base space involves complex polarization vectors.
Thus using a simple example of one-photon ionization we established that Eq.\eqref{eq:curvature_def} is associated with the orbital antisymmetric part of the molecular Berry curvature Eq.\eqref{eq:Omega_a}. Now we consider a general case to confirm these conclusions.

\section{4. Geometric Quantities Associated with Circular Polarization Vectors}
\begin{figure}
    \centering
    \includegraphics[width=0.4\linewidth]{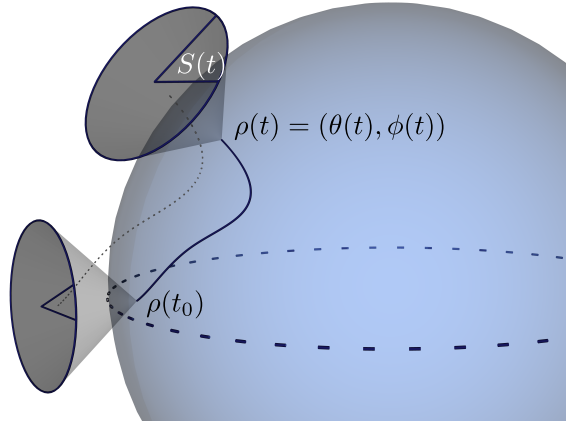} 
    \caption{The fiber bundle $\mathcal{H}\overset{\pi}{\rightarrow}S^2$. Each point in the projective base manifold (sphere) $S^2$. Each point on $S^2$ is a associated with a fiber of the the Hilbert space $\mathcal{H}$ which is determined by wavefunction $e^{iS}\ket{\psi}$. The angle $S$ parametrizes the fiber as can be seen for the two points $\rho(t_0)$ and $\rho(t)$. Throughout a time evolution a horizontal motion is given by the phase $S(t)$ (Eq.\,\eqref{eq: geometric phase}).}
    \label{fig:sphere_visualization}
\end{figure}
A generalized geometric picture arising from the dynamics of circular polarization vectors is given within the framework of fiber bundle theory. Such frameworks are often used in order to investigate quantum systems that either depend on an adiabatic parameter set \cite{berry1984quantal, cai1989covariant, nakahara2018geometry} or undergo a cyclic evolution \cite{aharonov1987phase, moore1990non, bouchiat1988non}. We partition the Hilbert space $\mathcal{H}$ of pure state, normalized wavefunctions $\ket{\psi(\bm{e})}$ into a set of fibers $\mathcal{H}_\rho = \left\{e^{iS}\ket{\psi(\bm{e}(\rho))}|S \in (-\pi,\pi]\right\}$ that consist of wavefunctions which are identical up to a $U(1)$-phase factor $e^{iS}$. By further defining the projection map $\pi(e^{iS}\ket{\psi(\bm{e}(\rho))}) = \frac{1}{2i}\bm{e}^*\times \bm{e}$ we obtain the unit sphere $S^2$ parametrized by the angles $\rho = (\theta,\phi)$ as projective space (or base manifold). 
The total construction of a total space, here the Hilbert space, endowed with a projective base manifold ($S^2$) and a projection map $\pi$ is referred to as the fiber bundle $\mathcal{H}\overset{\pi}{\rightarrow}S^2$. A schematic visualization is given in \Fig{\ref{fig:sphere_visualization}}. Here, for two points on the sphere $S^2$ the corresponding fibers $U(1)$ are visualized as circles. In a time evolution of the wavefunction $\ket{\psi(t)}$ the parameters $\rho(t)$ change as well as the phase $S(t)$. To this end, we distinguish between two kinds of dynamics in the total Hilbert space $\mathcal{H}$. A movement within one fiber, i.e. a change of phase, is referred to as vertical. (This motion is fully determined by the Lie group action of the unitary group $U(1)$ on the Hilbert space and, thus, extends the introduced bundle into a principal fiber bundle.) A notion of horizontal motion can be obtained by using the Berry connection form $\mathcal{A}$ \cite{nakahara2018geometry}. Throughout a time evolution the horizontal motion is given by the phase
\begin{align}\label{eq: geometric phase}
    S(t)= S(t_0) + i\int\limits_{\rho(t_0)}^{\rho(t)} \mathcal{A}.
\end{align}
The Berry connection form can be expressed in terms of the complex electric field polarization vector. Let $\nabla_{e_i}=\del{e_i}=\frac{1}{2}\left(\del{\mathrm{Re} e_i}-i\del{\mathrm{Im} e_i}\right)$ and $\nabla_{e_i^*}=\del{e_i^*} = \frac{1}{2}\left(\del{\mathrm{Re} e_i}+i\del{\mathrm{Im} e_i}\right)$ be the complex Wirtinger derivatives. In terms of the polarization vector, the Berry connection form reads \cite{wells_raymond, huybrechts2005complex} 
\begin{align} 
    \mathcal{A} &= i \braket{\psi|d \psi} = i \braket{\psi|\nabla_{\bm{e}} \psi} \cdot \bm{de}, 
\end{align} 
where the differentials of the polarization vector are $de_i = \bm{d\rho}  \cdot\nabla_{\rho} e_i = d\theta \del{\theta} e_i + d\phi \del{\phi} e_i$. Here we implicitly assumed the wavefunction to be a holomorphic section, i.e. $\ket{\nabla_{\bm{e}^*}\psi}=0$ \cite{ZuminoB}. This is in accordance with the perturbative approach of Section\,3. The Berry curvature form $\Omega$ is the exterior derivative of the Berry connection 
\begin{align} 
    \Omega 
    &= 
    d\mathcal{A} 
    = 
    i \braket{\partial_{e_i}\psi|\partial_{e_j}\psi} 
    de_i^* \wedge de_j. 
\end{align} 
Indeed, this expression is vectorizable in a way such that it nicely connects to physical observables. To this end, we define two bilinear vector products, 
\begin{align}
    (\bm{a} \diamond \bm{b})_i &\equiv |\varepsilon_{ilm}| a_l b_m, \hspace{60pt} (\bm{a} \odot \bm{b})_i \equiv \delta_{il} a_l b_l.
\end{align} 

The full Berry curvature then is 
\begin{align} 
    d\mathcal{A} 
    &= 
    i \bm{d\xi} \cdot \bra{\partial_{\bm{e}}\psi} \times \ket{\partial_{\bm{e}}\psi} 
    + i \bm{d\zeta} \cdot \bra{\partial_{\bm{e}}\psi} \diamond \ket{\partial_{\bm{e}}\psi} 
    + i \bm{d\chi} \cdot \bra{\partial_{\bm{e}}\psi} \odot \ket{\partial_{\bm{e}}\psi}, 
\end{align} 
where we defined the polarization differentials 
\begin{align} 
    d\zeta^l &\equiv \frac{1}{2}|\varepsilon^{lij}| de^{*i} \wedge de^{j}, \hspace{60pt} 
    d\chi^l \equiv \delta_{li} de^{*i} \wedge de^{i}. 
\end{align} 
All the above differentials have natural representations on the unit sphere in terms of the Euler angles $\rho$, 
\begin{align} 
    \bm{d\xi} 
    &= 
    d\theta \wedge d\phi \mathrm{Re}\left[\delc{\bm{e}^*}{\theta} \times \delc{\bm{e}}{\phi} \right], \\ 
    \bm{d\zeta} 
    &= 
    d\theta \wedge d\phi i \mathrm{Im} \left[ \delc{\bm{e}^*}{\theta} \diamond \delc{\bm{e}}{\phi} \right], \\ 
    \bm{d\chi} 
    &= 
    d\theta \wedge d\phi 2i \mathrm{Im} \left[ \delc{\bm{e}^*}{\theta} \odot \delc{\bm{e}}{\phi}\right]. 
\end{align} 
A detailed derivation of the Berry curvature are given  in Appendix\,A . 

\section{5. Berry Phase and Orbital Antisymmetric Berry Curvature in Two-Field Pump–Probe Experiments}
In the case of a pump-probe experiment, the two laser fields can have different polarization vectors. We summarize the geometric ingredients that arise when a pump–probe scheme is
driven by two laser fields with circular $\bm e= \frac{1}{\sqrt{2}}\left(\hat{\bm{\theta}}+i\sigma\hat{\bm{\phi}}\right)$ and a linear $\bm \varepsilon=\cos\alpha\,\hat{\bm{\theta}}+\sin\alpha\,\hat{\bm{\phi}}$ polarization vector. The angle $\alpha$ implies a phase shift between the fields. All derivational details are collected in Appendix B.\\

The Berry connection 1-form on the joint field manifold $(\bm e,\bm \varepsilon)$ is  
\begin{align}
\mathcal A =
i\,\langle\psi|\nabla_{(\bm e,\bm \varepsilon)}\psi\rangle \cdot d(\bm e,\bm \varepsilon),
\end{align}
which splits naturally into the circular‐ and linear-field contributions
\begin{align}
    \mathcal A =\mathcal A_{\bm e}+\mathcal A_{\bm \varepsilon}
\end{align}
and yields a geometric phase
\begin{align}
    S=\oint(\mathcal A_{\bm e}+\mathcal A_{\bm\varepsilon}).
\end{align}

Taking an exterior derivative gives the Berry curvature 2-form $\Omega = d\mathcal A$. Its orbital-antisymmetric part decomposes into
\begin{align}
    \Omega_a
   = \Omega_a^{\bm e}
     +\Omega_a^{\bm \varepsilon}
     +\Omega_a^{\text{cross}},    
\end{align}
with the compact vector formula 
\begin{align}
    \Omega_{a}^{\bm e} &= ( \nabla_{\bm{e}^*} \times \bm{A}_{\bm e} ) \cdot \bm{d\xi}_{\bm e} = \bm{\Omega}_{a}^{\bm e} \cdot \bm{d\xi}_{\bm e}
\end{align}
which coincides with the expression for the one-field orbital antisymmetric Berry curvature. The linear field polarization vector is real valued and thus its Berry curvature is fully antisymmetric:
\begin{align}
    \Omega^{\bm \varepsilon} = \Omega_{a}^{\bm \varepsilon} &= ( \nabla_{\bm \varepsilon} \times \bm{A}_{\bm \varepsilon} ) \cdot \bm{d\xi}_{\bm \varepsilon} = \bm{\Omega}_{a}^{\bm \varepsilon} \cdot \bm{d\xi}_{\bm \varepsilon} .
\end{align}
The mixed Berry curvature can be written as
\begin{align}
    \Omega^{\rm cross}_a &= - \text{Im}\left\{ \braket{\nabla_{\bm{\varepsilon}}\psi | \times | \nabla_{\bm e} \psi} \cdot \bm{d\xi}_{\rm cross} \right\} .
\end{align}
Neither the linear-field Berry curvature nor the mixed Berry curvature are relevant to isotropic signals (i.e. independent of the third Euler angle $\alpha$ distinguishing directions in polarization plane and breaking cylindrical symmetry) considered in our companion paper \cite{ordonez2024geometrytemporalchiralstructures}, and will  be analyzed in a separate publication, where all three Euler angles will be taken into account.

\section{6. Conclusion}
We have developed a unified geometric framework that highlights the interplay of geometric light–polarization dynamics and the photon-induced processes in chiral molecules. By treating the electronic wavefunction as a section of a $U(1)$ bundle over the space of molecular orientations, we can identify the fiber bundle connection and the orbital antisymmetric Berry curvature as the fundamental geometric objects governing enantiosensitive observables. By applying the formalism to one-photon ionization, we have shown that the antisymmetric part of the Berry curvature reduces to the net molecular propensity field, whose direction dictates the net orientation of the ionized ensemble \cite{ordonez2024geometrytemporalchiralstructures}.
In our companion paper we exploit the geometric framework of polarization vectors, where the Berry curvature describes a photoionization-driven enantio-sensitive  orientation of the molecular cations and the Berry connection governs the accumulation of geometric phases during rotational time evolutions. Taken together, the two studies demonstrate that the bundle curvature and connection provide a common language for describing both the enantiosensitive yield \cite{ordonez2023geometric} and the orientation of the cation. 
We have shown that the present formalism is readily generalized to multi-photon processes and pump-probe schemes. Since all observables follow from geometric quantities, the framework is conceptually transparent, making it a versatile tool for designing experiments and interpreting ultrafast chiral phenomena. We anticipate that further exploration of polarization-space geometry will provide control over singularities or non-trivial topologies that can be harnessed to amplify enantioselective signals.

\section{Acknowledgments}
We gratefully acknowledge many enlightening discussions with Prof. Misha Ivanov. We gratefully acknowledge Prof. Vladimir Chernyak for discussions and lectures on the topic.
 O.S., A.R., P.M.M. gratefully acknowledge ERC-2021-AdG project ULISSES, grant agreement No 101054696. Views and opinions expressed are however those of the authors only and do not necessarily reflect those of the European Union or the European Research Council. Neither the European Union nor the granting authority can be held responsible for them.
 
\section{Appendices} 
\subsection{Appendix A: Berry Curvature in Polarization Vector Space} \label{app: Curvature} 
\subsubsection{The Symmetrized Levi-Civita Symbol} 

\begin{lemma} 
\label{Lemma_eps}
    Let $\varepsilon$ be the Levi-Civita symbol in three dimensions. Then, 
    \begin{align} 
    |\varepsilon_{nij}||\varepsilon_{nlm}| 
    &= (1-\delta_{ij})(\delta_{il}\delta_{jm}+ \delta_{im}\delta_{jl}). 
    \end{align} 
\end{lemma} 
\textit{Proof:} Trivially, the equation holds true for $i=j$ and both sides of the equation equal zero. For the case $i\neq j$ we first consider $l=m$. The left-hand side of the equation equals zero and one the rhs we find 
    \begin{align} 
        \delta_{il}\delta_{jm}+ \delta_{im}\delta_{jl} 
        &= 
        \delta_{il}\delta_{jl}+ \delta_{il}\delta_{jl} \\ 
        &= 
        2\delta_{ij}\delta_{jl} \\ 
        &= 0. 
    \end{align} 
    The last case to prove is thus $i \neq j\cap l\neq m$. We rewrite $|\varepsilon_{nij}| = (1-\delta_{ki})(1-\delta_{ij})(1-\delta_{kj})$ such that 
    \begin{align} 
    |\varepsilon_{nij}||\varepsilon_{nlm}| 
    &= 
    \sum_k 
    (1-\delta_{ki})(1-\delta_{ij})(1-\delta_{kj}) 
    (1-\delta_{kl})(1-\delta_{lm})(1-\delta_{km}) \\ 
    &= 
    \sum_k 
    (1-\delta_{ki})(1-\delta_{kj}) 
    (1-\delta_{kl})(1-\delta_{km}) \\ 
    &= 
    \sum_k 
    (1-\delta_{ki}-\delta_{kj}+\delta_{ki}\delta_{kj})
    (1-\delta_{kl}-\delta_{km}+\delta_{kl}\delta_{km}) \\ 
    &= 
    \sum_k 
    (1-\delta_{ki}-\delta_{kj}+\delta_{ki}\delta_{kj}
    -\delta_{kl}+\delta_{kl}\delta_{ki}+\delta_{kl}\delta_{kj}-\delta_{kl}\delta_{ki}\delta_{kj} -\delta_{km}+\delta_{km}\delta_{ki} \\ 
    &+\delta_{km}\delta_{kj}-\delta_{km}\delta_{ki}\delta_{kj} 
    +\delta_{kl}\delta_{km}-\delta_{kl}\delta_{km}\delta_{ki}-\delta_{kl}\delta_{km}\delta_{kj}+\delta_{kl}\delta_{km}\delta_{ki}\delta_{kj}) 
     \\ 
    &= 
    -1+\delta_{ij}
+\delta_{il}+\delta_{jl}-\delta_{il}\delta_{ij}+\delta_{im} \\ 
    &+\delta_{jm}-\delta_{im}\delta_{ij}
    +\delta_{lm}-\delta_{il}\delta_{lm}-\delta_{jl}\delta_{lm}+\delta_{ij}\delta_{jl}\delta_{lm} 
     \\ 
     &= 
    -1 +\delta_{il}+\delta_{jl}+\delta_{im}+\delta_{jm} 
     \\ 
    \end{align} 
Thus, we need to prove 
\begin{align} 
    \delta_{il}\delta_{jm}+ \delta_{im}\delta_{jl} 
    &= 
    -1 +\delta_{il}+\delta_{jl}+\delta_{im}+\delta_{jm}, 
\end{align} 
for $i \neq j\cap l\neq m$. Since $i,j,l,m \in \{ 1,2,3 \}$ we have, again to distinct cases: $1. \,|\{i,j\}\cap \{l,m\}| =1 $ and $2. \,|\{i,j\}\cap \{l,m\}| =2 $. \\ 
\textbf{1. } 
W.l.o.g. we may assume that $i=l$ and $i,j,l\neq m$. Then, the equation amounts to $0=0$. \\ 
\textbf{2. } 
W.l.o.g. $i=l$ and $j=m$. Then, we have $1=1$. 

\subsubsection{Vectorial Expression for the Berry Curvature } 
The full Berry curvature is 
\begin{align} 
    \Omega 
    &= 
    i \braket{\partial_{e_i}\psi|\partial_{e_j}\psi} 
    de_i^* \wedge de_j. 
\end{align} 
We partition the tensor $i\braket{\partial_{e_i}\psi|\partial_{e_j}}$ into its symmetric and anti-symmetric components. For the symmetric components we separate the diagonal from the off-diagonal matrix elements, 
\begin{align} 
    \Omega 
    &= 
    \frac{i}{2}(\braket{\partial_{e_i}\psi|\partial_{e_j}\psi}-\braket{\partial_{e_j}\psi|\partial_{e_i}\psi}) 
    de_i^* \wedge de_j 
    + 
    \frac{i}{2}(\braket{\partial_{e_i}\psi|\partial_{e_j}\psi}+\braket{\partial_{e_j}\psi|\partial_{e_i}\psi}) 
    de_i^* \wedge de_j \\ 
    &= 
    \frac{i}{2}(\braket{\partial_{e_i}\psi|\partial_{e_j}\psi}-\braket{\partial_{e_j}\psi|\partial_{e_i}\psi}) 
    de_i^* \wedge de_j 
    + 
    \frac{i}{2}(1-\delta_{ij})(\braket{\partial_{e_i}\psi|\partial_{e_j}\psi}+\braket{\partial_{e_j}\psi|\partial_{e_i}\psi}) 
    de_i^* \wedge de_j \\ 
    &+ i\delta_{ij}\braket{\partial_{e_i}\psi|\partial_{e_j}\psi} 
    de_i^* \wedge de_j \\ 
    &= 
    \frac{i}{2}(\delta_{li}\delta_{mj}-\delta_{lj}\delta_{mi}) \braket{\partial_{e_l}\psi|\partial_{e_m}\psi} 
    de_i^* \wedge de_j 
    + 
    \frac{i}{2}(1-\delta_{ij})(\delta_{li}\delta_{mj}+ \delta_{lj}\delta_{mi}) \braket{\partial_{e_l}\psi|\partial_{e_m}\psi}  
    de_i^* \wedge de_j \\ 
    &+ i\delta_{ij}\braket{\partial_{e_i}\psi|\partial_{e_i}\psi} 
    de_j^* \wedge de_j. 
\end{align} 
We employ the identity $\varepsilon_{nij}\varepsilon_{nlm} = \delta_{il}\delta_{jm}+ \delta_{im}\delta_{jl}$ and Lemma\,\ref{Lemma_eps}, such that we may write 
\begin{align} 
    \Omega 
    &= 
    \frac{i}{2}\varepsilon_{nij}\varepsilon_{nlm} \braket{\partial_{e_l}\psi|\partial_{e_m}\psi} 
    de_i^* \wedge de_j 
    + 
    \frac{i}{2}|\varepsilon_{nij}||\varepsilon_{nlm}| \braket{\partial_{e_l}\psi|\partial_{e_m}\psi}  
    de_i^* \wedge de_j \\ 
    &+ i\delta_{ij}\braket{\partial_{e_i}\psi|\partial_{e_i}\psi} 
    de_j^* \wedge de_j. 
\end{align} 
We use the definitions of the standard cross product and the bi-linear forms ??? 
\begin{align} 
    \Omega 
    &= 
    i \bra{\partial_{\bm{e}}\psi}\times\ket{\partial_{\bm{e}}\psi} 
    \cdot \bm{d\xi} 
    + 
    i \bra{\partial_{\bm{e}}\psi}\diamond\ket{\partial_{\bm{e}}\psi} 
    \cdot \bm{d\zeta} 
    + i \bra{\partial_{\bm{e}}\psi}\odot\ket{\partial_{\bm{e}}\psi} 
    \cdot \bm{d\chi}. 
\end{align} 

\subsubsection{Euler-Angle Representation of the Polarization Differentials} 
For each of the differential two-forms $\bm{d\xi}$, $\bm{d\zeta}$ and $\bm{d\chi}$ an explicit expression in terms of the Euler angles $\rho=(\theta, \phi)$ is given by the exterior derivative of a one form $\omega$: $d\omega = d\rho_i \del{\rho_i} \omega$, 
\begin{align} 
    d\xi^l &= \frac{1}{2}\epsilon^{lij} \, de^{*i} \wedge de^j \\ 
    &= 
    \frac{1}{2}\epsilon^{lij} 
    \left(d\theta \del{\theta} + 
    d\phi \del{\phi} \right)e^{*i} \wedge \left(d\theta \del{\theta} + 
    d\phi \del{\phi} \right)e^j. 
\end{align} 
Due to the anti-symmetry of the wedge product we have $d\theta\wedge d\theta = d\phi\wedge d\phi = 0$ and $d\phi \wedge d\theta = - d\theta \wedge d\phi$, 
\begin{align} 
    d\xi^l &= 
    \frac{1}{2}d\theta \wedge d\phi \epsilon^{lij} 
    \left(\delc{e^{*i}}{\theta}\delc{e^j}{\phi} 
    - 
    \delc{e^j}{\theta}\delc{e^{*i}}{\phi}\right) \\ 
    &= 
    \frac{1}{2}d\theta \wedge d\phi 
    \left(\delc{\bm{e}^{*}}{\theta}\times\delc{\bm{e}}{\phi} 
    + 
    \delc{\bm{e}}{\theta}\times\delc{\bm{e}^{*}}{\phi}\right)^l \\ 
    &= 
    d\theta \wedge d\phi 
    \mathrm{Re}\left(\delc{\bm{e}^{*}}{\theta}\times\delc{\bm{e}}{\phi}\right)^l. 
\end{align} 
Similarly, we get 
\begin{align} 
    d\zeta^l &= \frac{1}{2}|\varepsilon^{lij}| de^{*i} \wedge de^{j} \\ 
    &= 
    \frac{1}{2}d\theta \wedge d\phi |\epsilon^{lij}| 
    \left(\delc{e^{*i}}{\theta}\delc{e^j}{\phi} 
    - 
    \delc{e^j}{\theta}\delc{e^{*i}}{\phi}\right) \\ 
    &= 
    \frac{1}{2}d\theta \wedge d\phi 
    \left(\delc{\bm{e}^{*}}{\theta}\diamond\delc{\bm{e}}{\phi} 
    - 
    \delc{\bm{e}}{\theta}\diamond\delc{\bm{e}^{*}}{\phi}\right)^l \\ 
    &= 
    d\theta \wedge d\phi 
    i\mathrm{Im}\left(\delc{\bm{e}^{*}}{\theta}\diamond\delc{\bm{e}}{\phi}\right)^l 
\end{align} 
and 
\begin{align} 
    d\chi^l &= \delta_{li} de^{*i} \wedge de^{i} \\ 
    &= 
    \delta_{li}d\theta \wedge d\phi 
    \left(\delc{e^{*i}}{\theta}\delc{e^i}{\phi} 
    - 
    \delc{e^i}{\theta}\delc{e^{*i}}{\phi}\right) \\ 
    &= 
    d\theta \wedge d\phi 
    \left(\delc{\bm{e}^{*}}{\theta}\odot\delc{\bm{e}}{\phi} 
    - 
    \delc{\bm{e}}{\theta}\odot\delc{\bm{e}^{*}}{\phi}\right)^l \\ 
    &= 
    d\theta \wedge d\phi 
    2i \mathrm{Im}\left(\delc{\bm{e}^{*}}{\theta}\odot\delc{\bm{e}}{\phi}\right)^l. \\ 
\end{align} 
\subsubsection{Berry Curvature for One-Photon Ionization} 
One-photon ionization processes are fully described by the first-order perturbative wave function 
\begin{align} 
\ket{\psi} & =\ket{0}+\int\mathrm{d}\Theta_{\bm{k}}a_{\bm{k}} (\bm{e})\ket{\bm{k}}. 
\end{align} 
The real, anti-symmetric part of the Berry curvature tensor $\Omega_{a}$ is determined by the vector  
\begin{align} 
    \bm{\Omega}_{a} &= \bra{\nabla_{\bm{e}}\psi}\times\ket{\nabla_{\bm{e}}\psi} \\ 
    &= \int\mathrm{d}\Theta_{\bm{k}} \int\mathrm{d}\Theta_{\bm{k}'}(\nabla_{\bm{e}^*}a_{\bm{k}}^*)\times(\nabla_{\bm{e}}a_{\bm{k}'}) \braket{\bm{k}|\bm{k}'} \\ 
    &= \int\mathrm{d}\Theta_{\bm{k}} (\nabla_{\bm{e}} a_{\bm{k}})^*\times(\nabla_{\bm{e}} a_{\bm{k}}). 
\end{align} 
The coefficient $a_{\bm{k}}$ may be expressed in terms of the ionization dipole $\bm{D} = \bra{\bm{k}}\hat{\bm{r}}\ket{0}$ and the circular field $\bm{E}(t)=E(t)\bm{e}$: $a_{\bm{k}}=-\widetilde{E}_{\omega_k}\bm{D}\cdot\bm{e}$ such that 
\begin{align}
    \bm{\Omega}_{a} &= 
    |\widetilde{E}_{\omega_k}|^2 \int\mathrm{d}\Theta_{\bm{k}} (\nabla_{\bm{e}} \bm{D}\cdot\bm{e})^*\times(\nabla_{\bm{e}} \bm{D}\cdot\bm{e}) \\ 
    &= 
    |\widetilde{E}_{\omega_k}|^2 \int\mathrm{d}\Theta_{\bm{k}} \bm{D}^*\times \bm{D}. 
\end{align} 
Analogously, we obtain 
\begin{align}
\bra{\nabla_{\bm{e}}\psi}\diamond\ket{\nabla_{\bm{e}}\psi} 
    &= 
    |\widetilde{E}_{\omega_k}|^2 \int\mathrm{d}\Theta_{\bm{k}} \bm{D}^*\diamond \bm{D}. 
\end{align} 
and 
\begin{align} 
\bra{\nabla_{\bm{e}}\psi}\odot\ket{\nabla_{\bm{e}}\psi} 
    &= 
    |\widetilde{E}_{\omega_k}|^2 \int\mathrm{d}\Theta_{\bm{k}} \bm{D}^*\odot \bm{D}. 
\end{align} 
Conclusively the full Berry curvature is 
\begin{align} 
    \Omega 
    &= 
    i |\widetilde{E}_{\omega_k}|^2 \int\mathrm{d}\Theta_{\bm{k}} \left( \bm{D}^*\times \bm{D} 
    \cdot \bm{d\xi} 
    + 
     \bm{D}^*\diamond \bm{D} 
    \cdot \bm{d\zeta} 
    + 
    \bm{D}^*\odot \bm{D} 
    \cdot \bm{d\chi} \right) \\ 
    &= 
    i d\theta \wedge d\phi 
    |\widetilde{E}_{\omega_k}|^2 \int\mathrm{d}\Theta_{\bm{k}} \\ 
    &\left( \bm{D}^*\times \bm{D} 
    \cdot \mathrm{Re}\left(\delc{\bm{e}^{*}}{\theta}\times\delc{\bm{e}}{\phi}\right) 
    + 
     \bm{D}^*\diamond \bm{D} 
    \cdot i\mathrm{Im}\left(\delc{\bm{e}^{*}} {\theta}\diamond\delc{\bm{e}}{\phi}\right) 
    + 
    \bm{D}^*\odot \bm{D} 
    \cdot 2i \mathrm{Im}\left(\delc{\bm{e}^{*}} {\theta}\odot\delc{\bm{e}}{\phi}\right) \right). 
\end{align}

\subsection{Appendix B: Berry Phase and Orbital Antisymmetric Berry Curvature in Two-Field Pump–Probe Experiments}
Here we derive the Berry connection and antisymmetric Berry curvature in the case of a two-field pump-probe Experiment.

\subsubsection{Berry Connection and Phase}
The Berry connection 1-form on the parameter manifold is
\begin{align}
    \mathcal{A} = i \braket{\psi | \nabla_{\rho}\psi} \cdot \mathrm{d}\rho\\
\end{align}
Since the wave function only depends on $\rho$ via the electric fields, we can express the gradient w.r.t. $\rho$ using the chain rule:
\begin{align}
\nabla_\rho
=
\frac{\partial\bm{e}}{\partial \rho}\,\nabla_{\bm{e}} + \frac{\partial\bm{\varepsilon}}{\partial \rho}\,\nabla_{\bm{\varepsilon}},
\end{align}
giving
\begin{align}
A = A_{e_i}\,de_i + A_{\epsilon_j}\,d\epsilon_j,
\end{align}
with
\begin{align}
A_{e_i} = i\,\bigl\langle \psi \,\bigl|\, \partial_{e_i}\,\psi \bigr\rangle,
\quad
A_{\epsilon_j} = i\,\bigl\langle \psi \,\bigl|\,\partial_{\epsilon_j}\,\psi \bigr\rangle.
\end{align}

The geometric (Berry) phase is defined as a loop integral in the parameter space~($\rho$):
\begin{equation}
    S 
    = i\oint \!\bra{\psi}\,\nabla_{\rho}\,\ket{\psi}\,\cdot\,\mathrm{d}\rho
    = i\oint {\bra{\psi}\,\nabla_{\bm{\varepsilon}}\,\ket{\psi}}\cdot d\bm{\varepsilon}
    +
    i\oint {\bra{\psi}\,\nabla_{\bm e}\,\ket{\psi}}\cdot\, d\bm{e}.
    \label{eq:S_as_sum_SE}
\end{equation}
Hence the Berry connection and phase and naturally split into two contributions, one for each field.

\subsubsection{Berry Curvature}
The Berry curvature 2--form is
\begin{align}
\Omega &= d\mathcal{A} \\
&= \partial_{e^*_i} A_{e_j} de^*_i \wedge de_j + \partial_{\epsilon_i} A_{e_j} d\epsilon_i \wedge de_j + \partial_{e^*_i} A_{\epsilon_j} de^*_i \wedge d\epsilon_j + \partial_{e_i} A_{\epsilon_j} de_i \wedge d\epsilon_j + \partial_{\epsilon_i} A_{\epsilon_j} d\epsilon_i \wedge d\epsilon_j\,,
\end{align}
where we assumed that $\psi$ is holomorphic in $\bm{e}$ and that $\bm{\varepsilon}$ is real.
Let's consider the orbital antisymmetric Berry curvature and identify 3 contributions:
\begin{align}
    \Omega_{a} &= \Omega_{a}^{\bm e} + \Omega_{a}^{\bm \varepsilon} + \Omega_{a}^{\rm cross}\\
    \intertext{with}
    \Omega_{a}^{\bm e} &= \frac12 \left(\partial_{e^*_i} A_{e_j} - \partial_{e^*_j} A_{e_i} \right) de^*_i \wedge de_j\\
    \Omega_{a}^{\bm \varepsilon} &= \frac12 \left( \partial_{\epsilon_i} A_{\epsilon_j} - \partial_{\epsilon_j} A_{\epsilon_i} \right) d\epsilon_i \wedge d\epsilon_j\\
    \Omega_{a}^{\rm cross} &= \frac12(\partial_{\epsilon_i} A_{e_j} - \partial_{\epsilon_j} A_{e_i}) d\epsilon_i \wedge de_j + \frac12( \partial_{e^*_i} A_{\epsilon_j} - \partial_{e^*_j} A_{\epsilon_i} ) de^*_i \wedge d\epsilon_j + \frac12( \partial_{e_i} A_{\epsilon_j} - \partial_{e_j} A_{\epsilon_i}) de_i \wedge d\epsilon_j
\end{align}
Next, we use the following identity for the contraction of a fully antisymmetric tensor $T_{ij}$ with some arbitrary tensor $K$, 
\begin{align}
    T_{ij} K_{ij} = \frac{1}{2} \epsilon_{nij} \epsilon_{nlm} T_{ij} K_{lm}
\end{align}
to rewrite the orbital antisymmetric Berry curvature:
\begin{align}
\Omega_{a}^{\bm e} &= \frac12 \bigl(\partial_{e^*_i} A_{e_j} - \partial_{e^*_j} A_{e_i}\bigr)\,de^*_i \wedge de_j\\
         &= \frac12 \epsilon_{nij} \bigl(\partial_{e^*_i} A_{e_j} - \partial_{e^*_j} A_{e_i}\bigr) \frac12 \epsilon_{nlm} de^*_l \wedge de_m .
\end{align}
Identifying $\bm{A}_{\bm e} = (A_{e_1},A_{e_2},A_{e_3})$ and $d\xi^{\bm e}_{k} = \frac12 \epsilon_{nlm} de^*_l \wedge de_m$ gives
\begin{align}
    \Omega_{a}^{\bm e} &= ( \nabla_{\bm{e}^*} \times \bm{A}_{\bm e} ) \cdot \bm{d\xi}_{\bm e} = \bm{\Omega}_{a}^{\bm e} \cdot \bm{d\xi}_{\bm e}
    \intertext{with}
    \bm{\Omega}_{a}^{\bm e} &= \nabla_{\bm{e}^*} \times \bm{A}_{\bm e} = i \braket{\nabla_{\bm e} \psi | \times | \nabla_{\bm e} \psi }
\end{align}
which coincides with the expression for the one-field orbital antisymmetric Berry curvature. Analogously, the linear field Berry curvature is fully antisymmetric:
\begin{align}
    \Omega^{\bm \varepsilon} = \Omega_{a}^{\bm \varepsilon} &= ( \nabla_{\bm \varepsilon} \times \bm{A}_{\bm \varepsilon} ) \cdot \bm{d\xi}_{\bm \varepsilon} = \bm{\Omega}_{a}^{\bm \varepsilon} \cdot \bm{d\xi}_{\bm \varepsilon},
    \intertext{with}
    \bm{\Omega}_{a}^{\bm \varepsilon} &= \nabla_{\bm \varepsilon} \times \bm{A}_{\bm \varepsilon} = i \braket{\nabla_{\bm \varepsilon} \psi | \times | \nabla_{\bm \varepsilon} \psi }.
\end{align}
The linear Berry curvature is equal to zero in one photon ionization.

\bibliography{Bibliography} 
\end{document}